\documentstyle[12pt]{article}
\begin{document}

\def\lsim{\:\raisebox{-0.5ex}{$\stackrel{\textstyle<}{\sim}$}\:}
\def\gsim{\:\raisebox{-0.5ex}{$\stackrel{\textstyle>}{\sim}$}\:}

\begin{flushright}
TIFR/TH/94-02 \\
\end{flushright}

\bigskip
\bigskip
\bigskip

\begin{center}
\large {\bf Slow Relaxation in a Model with Many Conservation Laws :
Deposition and Evaporation of Trimers on a Line} \\

\bigskip
\bigskip
\bigskip

Mustansir Barma and Deepak Dhar \\
Tata Institute of Fundamental Research \\
Homi Bhabha Road, Bombay 400 005, India \\

\bigskip
\bigskip
\bigskip

Abstract \\
\end{center}

\bigskip

\noindent We study the slow decay of the steady-state
autocorrelation function $C(t)$ in a
stochastic model of deposition and evaporation of trimers on a line with
infinitely many conservation laws and sectors.  Simulations show that
$C(t)$ decays as different powers of $t$, or as $\exp
(-t^{1/2})$, depending on the sector.  We explain this
diversity by relating the problem to diffusion of hard core particles
with conserved spin labels. The model embodies a matrix
generalization of the Kardar-Parisi-Zhang model of interface
roughening.  In the sector which includes the empty line, the dynamical
exponent $z$ is $2.55 \pm 0.15$.


\newpage

It is well known that conservation laws strongly affect the
decay in time of fluctuations in equilibrium.  For example, in magnetic
systems, the rate of decay of the spin autocorrelation function
is quite different depending on whether or not
the spin dynamics conserves magnetization \cite{HH}.
What happens with more than
one -- indeed an infinity of -- conservation laws?
We address this question in this paper by  studying the decay
of autocorrelation
functions in the steady state of a simple stochastic model of
deposition and evaporation \cite{BGS93,SGB93}. This model has been shown
to possess
an infinite number of independent constants of motion \cite{DB93}.
Here we show that it
exhibits a very rich variety of slow relaxations:
depending on the initial conditions, we can get a very large number of
power-law decays, or even stretched exponential decay.  We present
numerical evidence from Monte Carlo simulations for these different kinds
of decay, and also provide an analytical understanding of
this remarkable diversity in terms of the diffusive motion of hard core
particles with spin.  Our model can be viewed as a
generalisation of the
well-known Kardar-Parisi-Zhang (KPZ) problem \cite{KPZ} of roughening
of a one dimensional interface, where the scalar-height variables in the KPZ
model are replaced by a matrix-valued variables (say $2 \times 2$ complex
matrices).  In the steady state reached after starting from an
empty lattice, we find that the density-density autocorrelation function
decays as a power law in time, and the dynamical exponent $z \simeq 2.5$,
very different from the KPZ value $z =
3/2$.  The model thus includes a new universality class for
nonlinear one-dimensional evolution models.

In the deposition-evaporation (DE) model under consideration, there is
a site variable
$n_i$, taking values 0 or 1, at each site $i$ of a line, $1 \leq i \leq
L$; it may be thought of as an occupation number variable of a lattice
gas.  We define pseudospin variables $S_i \equiv 2n_i - 1$.  The time
evolution is Markovian, specified by the following rates: In a small time
$dt$, a triplet of spins at adjacent sites $i,i+1,i+2$
can flip simultaneously only if $S_i =
S_{i+1} = S_{i+2}$.  The rate is $\epsilon$ if the spins were originally
-1, and $\epsilon'$ if they were originally +1.
These rates satisfy the detailed balance condition corresponding to a
non-interacting lattice gas Hamiltonian with chemical potential ${1\over3}
\ell n(\epsilon/\epsilon')$.  However, the long-time steady states in the
present model are nontrivial, as the dynamics is strongly non-ergodic.  We have
shown \cite{DB93} that the total phase space of $2^L$ configurations
breaks up into a large number  $N_L$ of mutually
disconnected sectors, and
$N_L$ increases as $[(\sqrt{5} + 1)/2]^L$ for
large $L$.

This decomposition of phase space is a consequence of the
existence of an infinity
of conservation laws. A compact representation of these is provided by
the ``irreducible string''(IS), defined as follows: Regard the
configuration as a string of up and down spins.  Scan the string left to
right, and stop at the first triplet of parallel spins encountered.  If no
such triplet is found the string is the IS.  If
a triplet is encountered, simply delete it, reducing length of string by 3,
and repeat the process.  The IS which finally results is a
constant of motion, and different sectors of phase space correspond to
different IS's \cite{DB93}.

Under time evolution, an initial configuration evolves into the steady
state of the corresponding sector.  For instance, if $\epsilon =
\epsilon'$, in the steady state all configurations having the same IS as
that of the initial configuration occur with equal
probability.  It is then
interesting to ask how a dynamical quantity such as the steady state
autocorrelation function
$$
C_i (t) = \langle n_i (t+t_0) n_i (t_0)\rangle - \langle
n_i\rangle^2
\eqno (1)
$$
varies from one sector to another.  Figure 1 shows the behaviour of $C(t)$
obtained from Monte Carlo simulation studies
in a number of different representative sectors.
The data depicted in the figure pertain to four sectors in
which the IS constitutes a finite fraction of all sites, and one
sector in which the fraction is zero. These sectors correspond to
(1) a random initial configuration (2) an IS formed by repeating
$[\uparrow \downarrow \downarrow]$ $L/6$ times (3) an IS
formed by repeating
$[\uparrow \downarrow]$  $L/4$ times (4) an IS formed by repeating
$[\uparrow\uparrow\downarrow\uparrow\downarrow\downarrow]$ $L/12$
times (5) the null sector (vanishing IS).  Lattice sizes $L = 120,000$
were used and
averages over $30-60$ different histories were taken.  From the figure, we
note the following points: (i) if the lattice is divided into 3
sublattices $A,B,C$, the autocorrelation function $C_i (t)$ depends,
in general (e.g in Sectors 2 and 4),
on the sublattice to which site $i$ belongs.
(ii) in several sectors and sublattices $C_i (t)$ decays as a power law
$\sim t^{-\theta}$;  the power $\theta$ is sector dependent.  In
Sector 1, $\theta \simeq 0.25$; in Sector 2 and on two sublattices of
Sector 4, $\theta \simeq 0.5$;
in Sector 5, $\theta \simeq 0.6$.
(iii) in some
sectors (e.g. Sector 3 and one sublattice in Sector 4),
the decay is faster than a power law, and fits well to a
stretched exponential form.  Evidently, there is a very wide range of
possible behaviours.  Understanding the source and nature of this
dynamical diversity is one of the main points of this Letter.

Consider time evolution in a sector labelled by an IS with elements
$\{\alpha_n\} \equiv \{\alpha_1,\alpha_2,\cdots,\alpha_\ell\}$, where the
length $\ell$ is a nonzero fraction $\rho = \ell/L$ of the total length
$L$ of the lattice.
Let ${\cal C}(t)$ be the full string corresponding to the
configuration at time $t$.  If we apply the deletion
algorithm to ${\cal C}(t)$,
some characters are eventually deleted, others not.  If $X_j (t)$ is the
location of the $j^{\rm th}$ site (counting from the left)
that remains undeleted,  we may look
upon the set $\{X_j (t)\}$ as the positions of $\ell$
interacting random walkers
on a line; DE dynamics induces the time evolution
of$\{X_j (t)\}$. The walkers cannot cross each other
($X_{j+1} (t) > X_j
(t)$, for all $j$, all $t$) and they carry a conserved spin label
$S_{X_j(t)} = \alpha_j$ for all $t$.  Under time evolution $X_j (t)$
changes in jumps, by a multiple of 3 spaces at a time.

In the steady state, the joint probability distribution $Prob
(\{X_j\})$ is proportional to the number of different
configurations of the lattice, consistent with the positions of walkers
being $\{X_j\}$.  The substring between sites $X_{j+1}$ and $X_j$
should be a string reducible to a null string, and all such configurations
are easily enumerated.  The result is
$$
Prob(\{X_j\}) = {\cal N} \prod^{\ell}_{j=0} g(X_{j+1} - X_j)
\eqno (2)
$$
where ${\cal N}$ is a normalization constant,
and $X_0 = 0$, $X_{\ell+1} = L+1$.
The separation probability $g(r)$  can be computed using
the generating function method of \cite{DB93}. The result is
$g(r) \sim r^{-3/2} exp (-\kappa r)$ for large $r$, where
$\kappa$ is the reduced pressure,
which depends on on the density $\ell/L$, and tends to zero as $(\ell/L)^2$ for
small $\ell/L$.
We see that $\{X_j\}$ constitute the slow modes of the system
as they evolve diffusively, and are linked to
conserved quantities. If the typical relaxation time of a string of
length $\lambda$ reducible to the null string varies as $\lambda^z$,
then for times $t \gg (L/\ell)^z$ we can assume that all the degrees
of freedom other than $\{X_j\}$ are in  instantaneous local equilibrium.
Thus $g(r) \sim exp(-V(r))$,
where the effective interaction $V(r)$ between walkers is attractive, and
increases linearly with separation if $\kappa \ne 0$, and as $ln~r$ if
$\kappa = 0$.

While the  logarithmic part of the interaction is crucial
for the understanding of sectors in
which the walker density $\ell/L \rightarrow 0$, it appears not to be
very important if $\ell/L$, and hence
$\kappa$, is finite. In the latter case, we are
led to consider a simpler system: $\ell$ random walkers
on a line of length $L$, with each walker carrying a spin $\alpha_j$
which is unchanged under dynamics. Each
walker jumps left or right by  3
steps only if no other walkers intervene.  The
point is that this simpler system of hard core random walkers with spin
(HCRWS)
has the exactly the same conservation laws as the original deposition
evaporation model. Thus we are led to conjecture that
the long-time behaviour of $C_i(t)$ in a particular
sector of the DE model is
essentially the same
as the spin-spin autocorrelation function $D_i(t)$ in the HCRWS
problem with the corresponding spin sequence.

It is then easy to understand the sublattice dependence of $C_i (t)$ observed
in simulations.  As each element $\alpha_n$ of the IS moves by multiples
of 3 lattice spacings, it stays on the same sublattice, say $A$.  The
long-time behaviour of $C_i (t)$ for $i \in A$ is governed only by the
subset $\{\alpha_{n'}\}_A$ on sublattice $A$, independent of
$\{\alpha_{n'}\}_B$ and $\{\alpha_{n'}\}_C$, despite the
fact that the elementary step of
deposition or evaporation couples all sublattices strongly.

For the HCRWS model, mean squared fluctuations in the number of particles
between points 0 and $r$ grow as $t^{1/2}$ in the steady state
\cite{LIG}.  To
test for a similar behaviour in the DE system, we
monitored $\sigma^2 (t) \equiv \langle [N(r,t) - N(r,0)]^2\rangle$ where
$N(r,t)$ is the length of irreducible string between points 0 and $r$ at
time $t$.  This quantity is roughly equal to number of walkers to the left
of $r$, but remains well defined even in the sector where there are no
walkers $\ell = 0$.  Figure 2 shows Monte Carlo data for $\sigma^2 (t)$ in
a number of sectors.  For all sectors with $\ell/L$ finite,
$\sigma^2 (t)$ is seen to grow as $t^{1/2}$, lending strong
support to the conjecture of equivalence of the DE and HCRWS models.  In the
sector where the IS vanishes, on the other hand, $\sigma^2$ is found to grow
as $t^{2\beta}$ with $\beta \simeq 0.19$. If the irreducible string is very
short, then $\kappa$ is very small, and fluctuations in the separation
$(X_{i+1} - X_i)$ between near neighbours in the diffusing, interacting
lattice gas are large. The low value of the exponent $\beta$ implies
that these fluctuations are slowly decaying.

In the HCRWS model, the spin-spin autocorrelation function in the
steady state is defined as
$$
D_i(t)=\langle S_i (t+t_0) S_i(t_0) \rangle
- \langle S_i \rangle^2
\eqno (3)
$$
where $S_i (t)$ is spin at site $i$, and an unoccupied site is
considered to have zero spin.  $D_i (t)$ can be
written in the form
$$
D_i (t) = \sum^{+\infty}_{m=-\infty} G(m|t) \gamma_i (m)
\eqno (4)
$$
Here, given that in the steady state site $i$ occupied at some time $t_0$,
$G(m|t)$ is the conditional probability that it is occupied at
time $(t+t_0)$, and that the difference in the particle labels at
these times is $3m$.  This term is independent of the spin
configuration $\{\alpha_i\}$ of the IS.  The second term $\gamma_i
(m)$ is the average value of $\alpha_k \alpha_{k+3m}$, averaged over
different values of $k$ in the IS consistent with the condition that
particle $k$ can occupy site $i$ (i.e. $k=i$ mod 3).
The different dependences of $D_i(t)$ on time in different sectors
comes entirely from the different dependences of $\gamma_i(m)$ on $m$.

Using the known equivalence [7,8] of long-time HCRW dynamics to a
stochastic harmonic model, $G(m|t)$ for large $t$ is given by
$$
G(m|t) \approx {1 \over \sqrt{2 \pi\delta^2}} e^{-m^2/2\delta^2}
\eqno (5a)
$$
where
$$
\delta^2 =|m|erf \left({|m| \over \sqrt{2t}}\right) + \sqrt{2t \over \pi
} e^{-{m^2 \over 2t}}
\eqno (5b)
$$
Using eq. (4) and (5) we can determine the behaviour of $D_i (t)$ in
the HCRWS model in different sectors.

\noindent Sector 1 (random initial condition):  In this case $\gamma_i (m)$
is significant only for very small values of $m$.  For $m$ fixed, $G
\sim t^{-1/4}$ for large $t$, whence $D_i (t)$ also varies as
$t^{-1/4}$.  Note that this result is also true for other sectors in
which the correlations in the IS decay fast with distance.

\noindent Sector 2 $[\uparrow\downarrow\downarrow]$: In this case $\alpha_k
= \alpha_{k+3m}$ for all $m$.  Hence $D_i (t)$ reduces to a
density-density correlation which decreases as $t^{-1/2}$ for large
$t$.  The same behavior is expected in all periodic IS where
the magnitization in each sublattice is nonzero.

\noindent Sector 3 $[\uparrow\downarrow]$: In this case $\gamma_i (m) =
(-1)^m$, and this implies that $D_i (t)$ has the stretched exponential
form $\exp [-(t/\tau)^{1/2}]$ at large time at all sites.  The same
behaviour would occur whenever the IS is periodic with zero net
magnetization on each sublattice.

\noindent Sector 4
$[\uparrow\uparrow\downarrow\uparrow\downarrow\downarrow]$: On
sublattices $A$ and $C$ we have $\alpha_k = \alpha_{k+3m}$, and thus as
in sector 2, we have $D_i (t) \sim t^{-1/2}$.  On sublattice $B$,
$\gamma_i (m)$ is $(-1)^m$, and as in sector 3 we get stretched
exponential decay.

If the sector is such that correlations in the $IS$ imply that
$\tilde\gamma_i(q)$, the Fourier transform of $\gamma_i(m)$,
varies as $|q|^\phi$ for small $q$, then $D_i(t)$ would vary as
$t^{-(1+\phi)/4}$.

It is evident that the HCRWS predictions
for $D_i(t)$ accord very well with the Monte
Carlo results
for $C_i(t)$
shown in Fig. 1.  The occurrence of stretched exponentials
in the $[\uparrow\downarrow]$ sector, in particular, explains why earlier
attempts \cite{SGB93} to fit finite size data to power laws yielded
anomalously low values of the dynamical exponent $z$.

For any configuration $\{S_i\}$ of spins on the line, we define the
matrix variable at site $i$ by the equation
$$
I_i (t) = \prod^i_{j=1} A(S_j (t))
\eqno (6)
$$
where $A(1)$ and $A(-1)$ are $2 \times 2$ complex matrices given by
$$
A(1) = \left(\matrix{1 & x \cr 0 & \omega}\right) =
A(-1)^\dagger
\eqno (7)
$$
where $\omega \equiv exp(2 \pi i/3)$ and
$x$ is a real parameter.  Then as $A (1)^3 = A(-1)^3 = 1$, it
follows that $I_L (t)$ is a constant of motion \cite{DB93}.  The
evolution of the variables $\{I_i (t)\}$ is Markovian, governed by
local transition rates.  If we start with an empty line, $I_i (t=0)$
take simple values, and get progressively disordered as time
increases. For large $x$, the ratio $ln ~Tr~I_r(t)/ln~x$ behaves in
qualitatively the same way as $N(r,t)$. At $t=0$, the variance
$\sigma^2$ of $N(r,t)$ is zero,
and grows to a value
proportional to $r$ in the steady state.  If $N(r,t)$ obeyed a stochastic
evolution equation of the Kardar-Parisi-Zhang type \cite{KPZ}, this
variance would grow as $t^{2/3}$.  Our numerical results
(Fig. 2) clearly rule
out this possibility.  From Figs. 1 and 2 we see that
the decay of the autocorrelation function is characterized by $\theta
\simeq 0.6$, while correlations in the length of
irreducible string up to a given point grow anomalously
slowly, with $\beta \simeq 0.19$. Static correlation functions in this sector
are  characterized by power law decays as well.
For instance, if we define an indecomposable substring as
one which can be completely reduced, but which cannot be written as the
concatenation of smaller substrings reducible to
the null string \cite{DB93},
the probability of occurrence of an indecomposable substring  of length
$r$ varies as $r^{-3/2}$ for large $r$. This is
consistent with a random-walk
picture of  fluctuations in nearest-neighbour distance, and thus a
static correlation
exponent $\chi = 1/2$.  The dynamical exponent $z = \chi/\beta$ is thus
$\simeq 2.6$.  A numerical diagonalization study \cite{THD} of the spectrum of
the relaxation operator for finite systems with $L \leq 30$ yields a
value $2.55 \pm .15$. This value of $z$ is nonstandard, indicating that
the dynamics in this sector is characterized
by a new universality class.  In this sense, our
model is different then earlier attempt  \cite{DOH}
to generalize KPZ by allowing the height function to be an $N$ component
vector; it was argued that $z = 3/2$ independent of $N$ in $d=1$.

To summarize, we have shown that the deposition-evaporation model
exhibits great diversity
in the manner in which the autocorrelation function decays.  There is
strong evidence for the conjecture that in sectors with finite
IS density $\ell/L$, the dynamical behaviour is essentially the same
as that of the spin correlation function
of a system of hard core random walkers with the corresponding
spin sequence.
The diversity of relaxational behaviours in the DE model is
linked to differences in correlations in spin patterns in the IS
which labels that sector.
In the sector composed of completely reducible
configurations, the dynamics is characterized by a critical exponent
$z \simeq 2.5$, indicative of a new universality class.

MB acknowledges the hospitality of the Condensed Matter Group
at the International Center for Theoretical Physics, Trieste where
part of this work was done.

\newpage

\begin{center}
\underbar{\bf Figure Captions}
\end{center}
\bigskip

\noindent {\bf Figure 1:} Variation of the decay of the autocorrelation
function $C_i (t)$ in Sectors 1-5 (defined in text).  Sector
1 (open triangles); Sector 2, sublattice A (+), sublattices B and C
(filled triangles); Sector 3 (filled circles); Sector 4, sublattices
A and C (open squares), sublattice B (open circles); Sector 5
(filled squares).
The straight lines show power law decays $t^{-\theta}$ with
$\theta = 0.25$ (top), $0.50$ (middle), $0.62$ (bottom); the curve
shows a stretched exponential ($\sim \exp (-t^{1/2}))$ decay.

\bigskip
\bigskip

\noindent {\bf Figure 2:} Time-dependence of the mean squared fluctuations
in the length of irreducible string $N(r,t)$ up to point $r$.
The four sets of data
(from top) pertain to Sectors 1-4 (shifted vertically for clarity)
while the fifth pertains to Sector 5.  The straight lines have slopes
0.50 and 0.38.

\end{document}